# Spectral solution of axisymmetric magnetization problems for thin superconducting shells

Leonid Prigozhin and Vladimir Sokolovsky

*Abstract*—Existing numerical methods for modeling magnetization in thin type-II superconducting films have mostly been developed for flat films. This work introduces an efficient spectral method for non-flat axisymmetric film magnetization problems. The method is based on the integral thin-shell current-density formulation of the problem, employs Chebyshev polynomial expansions for spatial discretization, and uses the method of lines for time integration. It applies to both open and closed axisymmetric shells and is so accurate that the solutions obtained can serve as benchmarks for numerical methods for general, not necessarily axisymmetric, thin-shell magnetization problems. As one of the examples, we consider magnetic shielding by a superconducting sphere.

## I. Introduction

THE finite element method is the primary tool for solving problems in applied superconductivity and, starting from [1-3], a variety of such methods have been developed to model magnetization in thin superconducting films. With a few exceptions (e.g., strips in cables [4] and a conical film [5]), these methods focus on flat films rather than superconducting shells of arbitrary shape. Furthermore, for most of these methods, it can be challenging to determine the electric field and then estimate AC losses or voltages generated by magnetic pumps. Indeed, the vectorial sheet current density, if its direction is unknown in advance, is usually replaced in thin-film problems by its scalar stream function (T-potential) and then calculated as the curl of this function. Although the stream function can be calculated with high accuracy, numerical differentiation reduces the precision of the computed current density. The highly nonlinear current-voltage relation of the superconducting material, when used directly to determine the electric field, results in much higher numerical errors and renders the field calculation unreliable. A similar difficulty arises in the FFT-based method [6, 7]. To address this issue, two mixed finite element methods based on the T-E formulation for flat-film magnetization problems have been developed in [8] and [9]. We have now extended the method [9] to non-flat films [10]. The known analytical solutions for flat films [11, 12] would be insufficient for verifying the new method.

Spectral methods, which seek solutions in the form of, for example, a Fourier series or a Chebyshev polynomial expansion, are less versatile than finite elements: their use can be difficult for problems with complex geometry. However, if applicable, these methods converge to a smooth solution exponentially, much faster than finite elements. Recently, several problems for superconducting strips, stacks, and coils were solved using spectral spatial discretization on their integral formulations [13-18]. Here, we extend this approach to axisymmetric thin-shell magnetization problems. This enables us to generate a wide variety of solutions to non-flat shell problems with both the sheet current density and the electric field so precise that they can serve as benchmark references, like analytical solutions.

As an example, we consider magnetic shielding by a superconducting shell. For the Meissner state of a thin-wall superconducting sphere (with or without a hole) or a cylinder (with or without cups), axisymmetric shielding problems have been studied in, e.g., [19-22]. Various finite element methods have been employed to model shielding by bulk type-II superconductors in the mixed state (see the recent review [23] and the references therein). Our method automatically detects transitions between the ideal Meissner state and the mixed state of a superconducting shell caused by changes in the external field.

## II. Integral Formulation of the Magnetization Problem

We assume the thickness of an axisymmetric, type-II superconducting shell is much smaller than its other dimensions and use the thin-shell model. In this model, the shell is represented by its mid-surface $S$ with a nonlinear relation between the tangential electric field $E$ and the surface (sheet) current density $J$.

Let $S$ be generated by rotation around the $z$ axis of a smooth, non-self-crossing curve parametrically defined in

L. Prigozhin is with Blaustein Institutes for Desert Research, Ben-Gurion University of the Negev, Sde Boqer Campus 84990, Israel (e-mail: leonid@bgu.ac.il); corresponding author.

V. Sokolovsky is with Physics Department, Ben-Gurion University of the Negev, Beer-Sheva 84105 Israel (e-mail: sokolovv@bgu.ac.il)
Color versions of the figures in this article are available online at http://ieeexplore.ieee.org

cylindrical coordinates $(r,\phi,z)$ as $\varsigma = \{r = R(s), z = Z(s)\}$ for $\phi = 0$. Here $-1 \leq s \leq 1$ and we assume $R(s) > 0$ for $s \neq \pm 1$. The surface is closed (e.g., a torus or a sphere) if the curve $\varsigma$ is closed or if $R(-1) = R(1) = 0$; otherwise, it has one or two boundaries (e.g., a conical or a cylindrical shell). The sheet current density, induced by the time variations of an axisymmetric external field $\boldsymbol{H}^e$, has only the azimuthal component, $\boldsymbol{J} = J(t,s)\boldsymbol{i}_\phi$, and its vector potential $\boldsymbol{A}[\boldsymbol{J}]$ can be chosen as $A_\phi \boldsymbol{i}_\phi$. On the surface $S$ we have

$$A_\phi(t,s) = \frac{\mu_0}{4\pi} \int_{-1}^{1} \int_{0}^{2\pi} \frac{J(t,s')\cos\phi'}{D(s,s',\phi')} R(s')\chi(s') d\phi' ds',$$

where $\chi = \sqrt{(dR/ds)^2 + (dZ/ds)^2}$, $R(s)\chi(s)d\phi ds$ is the surface element $dS$, and

$$D = \sqrt{R(s)^2 + R(s')^2 + (Z(s) - Z(s'))^2 - 2R(s)R(s')\cos(\phi')}.$$

Denoting

$$m = \frac{4R(s)R(s')}{(R(s) + R(s'))^2 + (Z(s) - Z(s'))^2}$$

and using the known expression for the integral over $\phi$ via the elliptic integrals $K$ and $E$,

$$\int_{0}^{2\pi} \frac{\cos\phi'}{D(s,s',\phi')} d\phi' = \frac{4}{\sqrt{mR(s)R(s')}} \left[(1 - m/2)K(m) - E(m)\right],$$

we obtain

$$A_\phi(t,s) = \frac{\mu_0}{\pi} \int_{-1}^{1} \mathcal{K}(s,s')\chi(s')J(t,s')ds'$$

with

$$\mathcal{K}(s,s') = \frac{1}{\sqrt{m}} \sqrt{\frac{R(s)}{R(s')}} \left[(1 - m/2)K(m) - E(m)\right].$$

The vector potential of the axisymmetric external field can be chosen as $\boldsymbol{A}^e = A_\phi^e \boldsymbol{i}_\phi$, too. For a uniform field $\boldsymbol{H}^e = H_z^e(t)\boldsymbol{i}_z$ we set $\boldsymbol{A}^e = \left[\mu_0 r H_z^e(t)/2\right]\boldsymbol{i}_\phi$.

In this problem, the variables $\boldsymbol{J}$ and $\boldsymbol{E}$ have the same (azimuthal) direction, so we can regard them as scalars. We assume $E = \rho J$ with the power law surface resistivity

$$\rho = E_c / J_c \left(|J|/J_c\right)^{n-1}, \quad (1)$$

where the critical field $E_c$, power $n$, and the critical sheet current density $J_c$ are constants. Expressing the tangential electric field component via the vector and scalar potentials, we obtain $(E + \partial_t(A_\phi[J] + A_\phi^e)\boldsymbol{i}_\phi = -[\nabla\Phi]_\tau$. To satisfy this equation, $[\nabla\Phi]_\tau$ should, as is true for the other terms of this equation, be parallel to $\boldsymbol{i}_\phi$ and cannot depend on $\varphi$. Hence, $\partial_\phi \Phi = C(t,s)$ and, integrating, we find $\Phi|_{\phi=2\pi} - \Phi|_{\phi=0} = 2\pi C(t,r,z)$, which means $C = 0$. The axisymmetric thin shell magnetization problem can now be written as the evolutionary integro-differential equation

$$E(J(t,s)) + \frac{\mu_0}{\pi} \partial_t \int_{-1}^{1} \mathcal{K}(s,s')\chi(s')J(t,s')ds' = -\partial_t A_\phi^e(t,s) \quad (2)$$

supplemented by the initial condition $J|_{t=0} = J_0(s)$. An advantage of this formulation is that it is spatially one-dimensional and is written exclusively in terms of the sheet current density: no outer space consideration is necessary.

Provided the sheet current density is determined, the magnetic field in the surrounding space can be computed by integrating the known expressions for the field of a circular current loop,

$$H_z = \int_{-1}^{1} K_z(r, R(s), z - Z(s))\chi(s)J(s)ds + H_z^e(r,z),$$

$$H_r = \int_{-1}^{1} K_r(r, R(s), z - Z(s))\chi(s)J(s)ds + H_r^e(r,z), \quad (3)$$

where

$$K_z(r,R,z) = \frac{K(\hat{m}) + \frac{R^2 - r^2 - z^2}{(R-r)^2 + z^2} E(\hat{m})}{2\pi\sqrt{(R+r)^2 + z^2}},$$

$$K_r(r,R,z) = \frac{z\left[-K(\hat{m}) + \frac{R^2 + r^2 + z^2}{(R-r)^2 + z^2} E(\hat{m})\right]}{2\pi\sqrt{(R+r)^2 + z^2}}$$

with $\hat{m} = 4rR/\left[(R+r)^2 + z^2\right]$.

III. CHEBYSHEV SPECTRAL METHOD

The kernel $\mathcal{K}(s,s')\chi(s')$ of the integral operator in (2) is singular because the elliptic integral $K(m)$ tends to infinity as $m \to 1$. Separating its singular part (see Appendix), we rewrite (2) as

$$E(J(t,s)) + \frac{\mu_0}{\pi} \times$$

$$\partial_t \left[ \int_{-1}^{1} \mathcal{U}(s,s') J(t,s') ds' - \frac{\chi(s)}{2} \int_{-1}^{1} \ln|s-s'| J(t,s') ds' \right] \quad (4)$$

$$= -\partial_t A_\phi^e(t,s),$$

where the kernel $\mathcal{U} = \mathcal{K}(s,s')\chi(s') + (\chi(s)/2)\ln|s-s'|$ is a regular function. Chebyshev polynomial expansions are known to be a convenient tool for solving the integral equations with logarithmic singularities (see, e.g., [24]). We use interpolating expansions in Chebyshev polynomials of the first kind, $T_k(s)$, and we will now briefly describe their properties (see [25]). By definition, $T_k(s) = \cos(k \arccos(s))$ for $s \in [-1,1]$. It is a polynomial of degree $k$ and its values can be computed recursively:

$$T_0(s) = 1, \quad T_1(s) = s,$$
$$T_k(s) = 2s T_{k-1}(s) - T_{k-2}(s) \text{ for } k = 2,3,... \quad (5)$$

These polynomials are orthogonal in $[-1,1]$ with respect to the weight $(1-s^2)^{-1/2}$:

$$\int_{-1}^{1} \frac{T_i(s) T_j(s)}{\sqrt{1-s^2}} ds = \begin{cases} 0 & i \neq j, \\ \pi & i = j = 0, \\ \pi/2 & i = j > 0. \end{cases} \quad (6)$$

Let $\{s_1, s_2, ..., s_N\}$ be a set of non-coinciding points in $[-1,1]$, $\overline{f} = (f(s_1), ..., f(s_N))^T$ the vector of values of a function $f$ at these points, and the expansion in Chebyshev polynomials $\sum_{j=0}^{N-1} \eta_j T_j(s)$ interpolate $f$ at the points $s_k$. Then $\overline{f} = P\overline{\eta}$, where $P_{ij} = T_j(s_i)$, and $\overline{\eta} = P^{-1}\overline{f}$. Recursive relations (5) can be used to efficiently compute the matrix $P$. We use polynomial expansions that interpolate functions at the Chebyshev points of the first kind,

$$s_k = \cos(\pi(k-1/2)/N), \quad k = 1,...,N.$$

These points are denser near the interval ends, which suppresses the Runge phenomenon, the wild oscillations of polynomials of high degrees between the interpolation knots.

Let us denote $g_k(t) = J(t, s_k)$ and seek the approximate solution to (4) in the form of a weighted interpolating expansion, convenient for computing the integral terms of this equation:

$$J(t,s) \approx G(t,s) = \frac{\sum_{j=0}^{N-1} \alpha_j(t) T_j(s)}{\sqrt{1-s^2}}. \quad (7)$$

The expansion coefficients $\overline{\alpha} = P^{-1} Y \overline{g}$, where $Y$ is the diagonal matrix with $Y_{kk} = \sqrt{1-s_k^2}$. The coefficients of the bivariate interpolating expansion $\sum_{i,j=0}^{N-1} \beta_{ij} T_i(s) T_j(s')$ of $\mathcal{U}(s,s')$ are elements of the matrix $\beta = P^{-1} U (P^{-1})^T$, where $U_{ij} = \mathcal{U}(s_i, s_j)$. Using (6), (7), and denoting $\sigma_0 = \pi$, $\sigma_{j>0} = \pi/2$, we obtain:

$$\int_{-1}^{1} \mathcal{U}(s,s') \partial_t G(t,s') ds' = \sum_{j=0}^{N-1} \dot{\alpha}_j(t) \int_{-1}^{1} \frac{\mathcal{U}(s,s') T_j(s')}{\sqrt{1-(s')^2}} ds' =$$

$$\sum_{j=0}^{N-1} \sigma_j \dot{\alpha}_j(t) \sum_{i=0}^{N-1} \beta_{ij} T_i(s) = \sum_{i=0}^{N-1} \left[ \sum_{j=0}^{N-1} \beta_{ij} \sigma_j \dot{\alpha}_j(t) \right] T_i(s) = \sum_{i=0}^{N-1} \gamma_i T_i(s),$$

where $\dot{\alpha}_j$ is the time derivative of $\alpha_j$. Introducing the diagonal $N \times N$ matrix $\Sigma$ with $\Sigma_{kk} = \sigma_{k-1}$, we rewrite this result in matrix form, $\overline{\gamma} = \beta \Sigma \dot{\overline{\alpha}} = \beta \Sigma P^{-1} Y \dot{\overline{g}} = V \dot{\overline{g}}$.

We present $\ln|s-s'|$ by its known Chebyshev expansion [24],

$$\ln|s-s'| = -\ln 2 - \sum_{j=1}^{\infty} \frac{2}{j} T_j(s) T_j(s') = -\sum_{j=0}^{\infty} \lambda_j T_j(s) T_j(s').$$

This yields

$$\Psi(t,s) = \int_{-1}^{1} \ln|s-s'| \partial_t G(t,s') ds' =$$

$$\sum_{j=0}^{N-1} \dot{\alpha}_j(t) \int_{-1}^{1} \frac{\ln|s-s'| T_j(s')}{\sqrt{1-(s')^2}} ds' =$$

$$-\sum_{j=0}^{N-1} \dot{\alpha}_j(t) \lambda_j \sigma_j T_j(s) = -\sum_{j=0}^{N-1} \kappa_j T_j(s),$$

where the vector of coefficients $\overline{\kappa} = \Lambda \Sigma \dot{\overline{\alpha}} = \Lambda \Sigma P^{-1} Y \dot{\overline{g}}$ and $\Lambda$ is the diagonal matrix with $\Lambda_{kk} = \lambda_{k-1}$. To find the interpolating expansion of the term $\chi(s)\Psi(t,s)/2$ in (4), we multiply the vector $\overline{\psi} = -P\overline{\kappa}$ of $\Psi(t,s)$ values at the points $s_k$ by the diagonal matrix $Q$ with $Q_{kk} = \chi(s_k)/2$ and find the expansion coefficients:

$$\frac{\chi(s)}{2} \int_{-1}^{1} \ln|s-s'| \partial_t G(t,s') ds' = \sum_{j=0}^{N-1} \delta_j(t) T_j(s)$$

where
$$\overline{\delta} = P^{-1} Q \overline{\psi} = -P^{-1} Q P \overline{\kappa} = -P^{-1} Q P \Lambda \Sigma P^{-1} Y \dot{\overline{g}} = -W \dot{\overline{g}}.$$

The interpolating expansion of $E(J)$ has the coefficients $P^{-1}\overline{e}$, where $e_k = E(g_k)$. We also find the expansion coefficients, $P^{-1}\overline{f}^e$, of the right-hand side $f^e = -\partial_t A^e_\phi(t,s)$. Equating the expansion coefficients of the left and right sides of (4), we arrive at the ordinary differential equation (ODE) system $\frac{\mu_0}{\pi}(V+W)\dot{\overline{g}} = P^{-1}(\overline{f}^e - \overline{e})$, or $\dot{\overline{g}} = M(\overline{f}^e - \overline{e})$ with $M = \frac{\pi}{\mu_0}(V+W)^{-1}P^{-1}$. This system can be solved numerically by a standard ODE solver.

Finally, to find the magnetic field outside the shell, the integrals in (3) can be computed numerically using the Gauss-Chebyshev quadrature

$$\int_{-1}^{1} u(s)\mathrm{d}s \approx \frac{\pi}{N}\sum_{i=1}^{N} u(s_i)\sqrt{1-s_i^2}. \quad (8)$$

IV. NUMERICAL SIMULATION RESULTS

Our simulations were performed in MATLAB R2020b on a PC with the Intel i7-10700 CPU and 64 GB RAM. The MATLAB ODE solver ode15s was used with both the relative and absolute tolerances set to $10^{-9}$. We present the results using scaled dimensionless variables

$$\tilde{J} = \frac{J}{J_c}, \quad \tilde{H} = \frac{H}{J_c}, \quad \tilde{E} = \frac{E}{E_c}, \quad (\tilde{r}, \tilde{z}) = \frac{(r,z)}{a}, \quad \tilde{t} = \frac{t}{t_0},$$

where $a$ is the characteristic shell size and $t_0 = a\mu_0 J_c / E_c$. In all examples, we assume $\tilde{J}(0,s) = 0$ and the external field is uniform, $\tilde{\boldsymbol{H}}^e = \tilde{H}^e_z(t)\boldsymbol{i}_z$, with $\tilde{H}^e_z = k_H \tilde{t}$. We set the field growth rate $k_H = 6$; this rate primarily affects the magnitude of the electric field.

Let us start with a thin disk and compare our solution with the known analytical one for the Bean critical-state model: the sheet current density, found in [12], is

$$\tilde{J} = \begin{cases} -1 & b \le \tilde{r} \le 1, \\ -\frac{2}{\pi}\arctan\left(\tilde{r}\sqrt{\frac{1-b^2}{b^2-\tilde{r}^2}}\right) & 0 \le \tilde{r} < b, \end{cases} \quad (9)$$

where $b(t) = 1/\cosh\left[2\tilde{H}^e_z(t)\right]$. The multivalued $E(J)$ relation of the Bean model does not allow one to find the corresponding electric field directly. We calculated it numerically, as in [8], computing the normal-to-film magnetic field $\tilde{H}_z = \tilde{H}_z[\tilde{J}] + \tilde{H}^e_z$ at $\tilde{t} \pm \varepsilon$ for a small $\varepsilon > 0$ and integrating the equation

$$\tilde{r}^{-1}\partial_r(\tilde{r}\tilde{E}) = -(\tilde{H}_z|_{\tilde{t}+\varepsilon} - \tilde{H}_z|_{\tilde{t}-\varepsilon})/(2\varepsilon)$$

with $\tilde{E}|_{\tilde{r}=0} = 0$.

The Bean model solution (9) is also the $n \to \infty$ limit of solutions to models with the power law (1), see [26]. Setting $\varsigma = \{\tilde{r} = (s+1)/2, \tilde{z} = 0\}$, we studied the convergence as both the power $n$ and the number of grid points $N$ increase. Since the solution (9) is not smooth (its first derivative is discontinuous), we cannot expect a very fast convergence of the spectral method for large values of $n$. Nevertheless, the solutions obtained are close (Fig. 1). In this example, we characterized the deviation from the Bean model solution by the relative errors $\delta J$, $\delta E$ in the integral $L^2$-norm approximated using (8), see Table I.

**Fig. 1.** Thin disk in the field $\tilde{H}^e_z = 6\tilde{t}$ at $\tilde{t} = 0.1$. The Bean

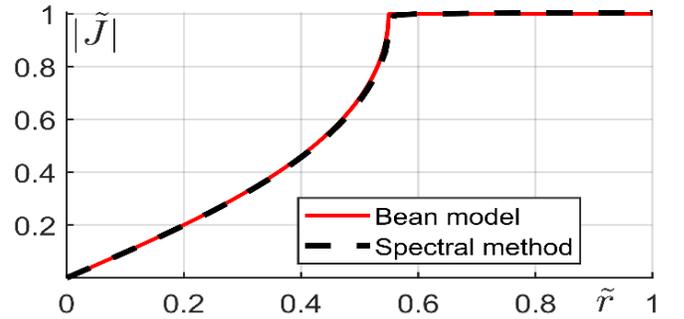

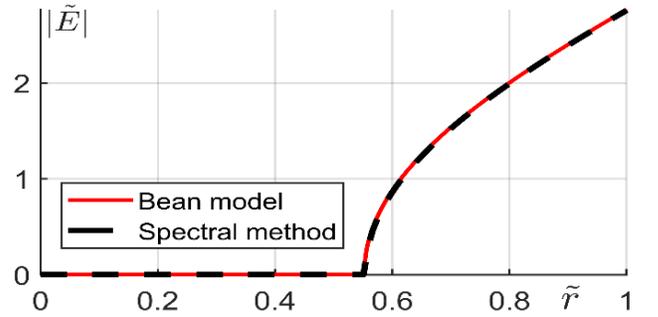

model solution and the spectral solution with $n = 200$, $N = 200$.

TABLE I

| Thin disk, $\tilde{H}^e_z = 6\tilde{t}, \tilde{t} = 0.1$. Relative deviations of spectral solutions from the Bean model (in L²-norm) | | | | |
|---|---|---|---|---|
| $n$ | $N$ | $\delta J$, % | $\delta E$, % | CPU time, sec. |
| 50 | 50 | 1.7 | 1.9 | 0.15 |
| 200 | 200 | 0.9 | 0.7 | 3.6 |

In other examples, we assumed a more realistic power value, $n = 20$. The solutions are now smoother, and we investigate the convergence rate of the proposed method by computing the maximal absolute deviations $\Delta\tilde{J}, \Delta\tilde{E}$ of numerical solutions from the most accurate one. To compare solutions obtained with different Chebyshev meshes, they are transformed first into Chebyshev interpolating expansions, then the expansion values at the fixed uniform grid of 100 points in $(-1,1)$ are computed and compared.

For a closed spherical shell, we define $\varsigma = \{\tilde{r} = \cos(\pi s/2), \tilde{z} = \sin(\pi s/2)\}$. While the external field is weak, the sheet current density is everywhere subcritical, $|\tilde{J}|<1$, and the electric field $\tilde{E}$ is practically zero, which corresponds to the whole shell being in the Meissner state with an almost ideally shielded interior. As the field $\tilde{H}_z^e$ continues to grow, there appears an overcritical sheet current density region, where $|\tilde{J}|>1$, the electric field is non-zero, and through which the field penetrates inside the sphere (Figs. 2-4). At the sphere center, the magnetic field remains zero for $|\tilde{H}^e|$ less than, approximately, $0.65$; at $|\tilde{H}^e|=1$ the shielding coefficient is about 5 (Fig. 4). To investigate the convergence, we performed simulations with different meshes and compared the results at $\tilde{H}_z^e = 0.9$ with those for the finest mesh, $N = 800$, see Table II.

Magnetization of another closed shell, a torus with the generator $\varsigma = \{\tilde{r} = 2 + \cos(\pi s), \tilde{z} = \sin(\pi s)\}$, is illustrated in Fig. 5. See also the convergence results for the solution at $\tilde{H}_z^e = 0.9$ in Table III.

The two more examples (Fig. 6) are for a cylindrical shell, $\varsigma = \{\tilde{r} = 1, \tilde{z} = s\}$, and a cylindrical shell closed by a hemisphere from below. In the latter case, we set

$$\varsigma = \left\{\tilde{r} = \begin{bmatrix} \sin(l) & l < \pi/2 \\ 1 & l \geq \pi/2 \end{bmatrix}, \tilde{z} = \begin{bmatrix} -\cos(l) & l < \pi/2 \\ l - \pi/2 & l \geq \pi/2 \end{bmatrix}\right\},$$

with $l(s) = a(s+1)$, where $a = \pi/4 + 1/2$ is the half-length of the generator curve. Convergence results for the semi-closed cylinder example are presented in Table IV for the external field $\tilde{H}_z^e = 0.7$.

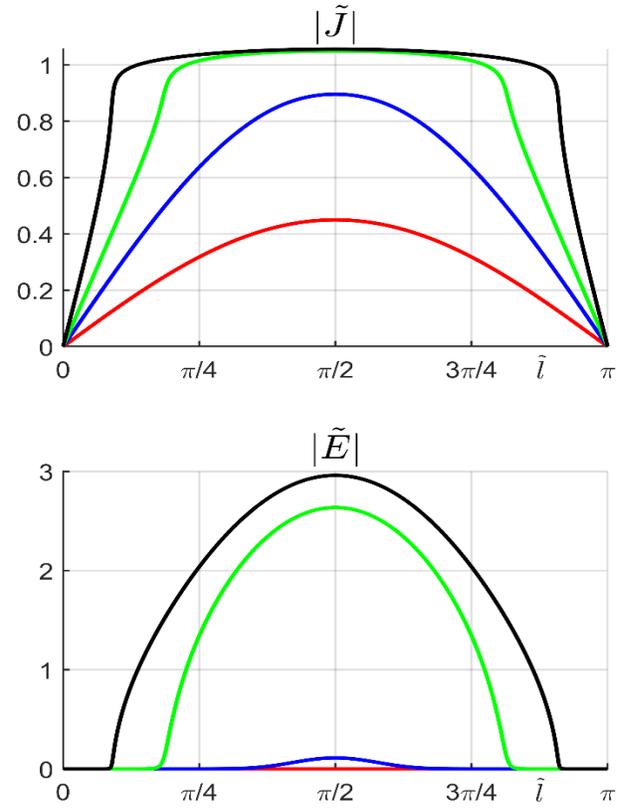

**Fig. 2.** Distributions of the sheet current density (top) and electric field (bottom) for $n = 20$, $\tilde{H}_z^e = 0.3$ (red), 0.6 (blue), 0.9 (green), 1.2 (black). $N = 400$; $\tilde{l}$ is the arc length.

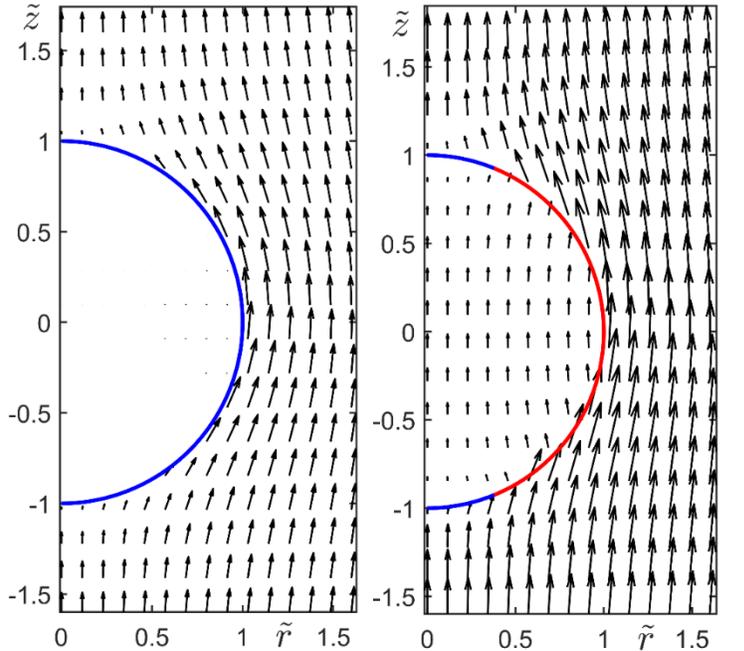

**Fig. 3.** Magnetic field in the vicinity of a superconducting sphere, $\tilde{H}_z^e = 0.6$ (left) and 1.2 (right). Red line indicates the $|\tilde{J}|>1$ zone.

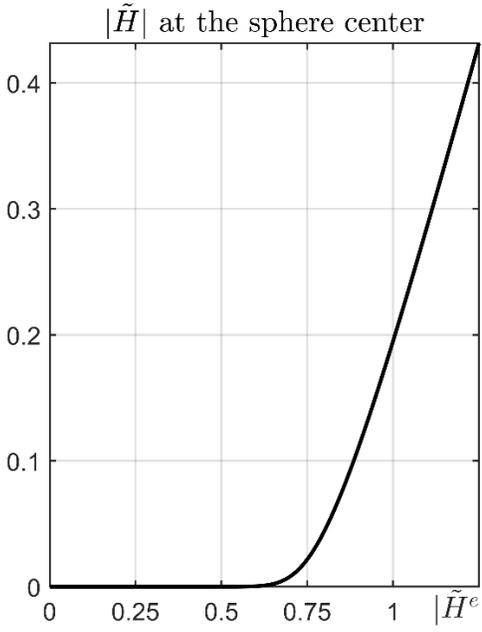

**Fig. 4.** Magnetic field at the sphere center vs external field.

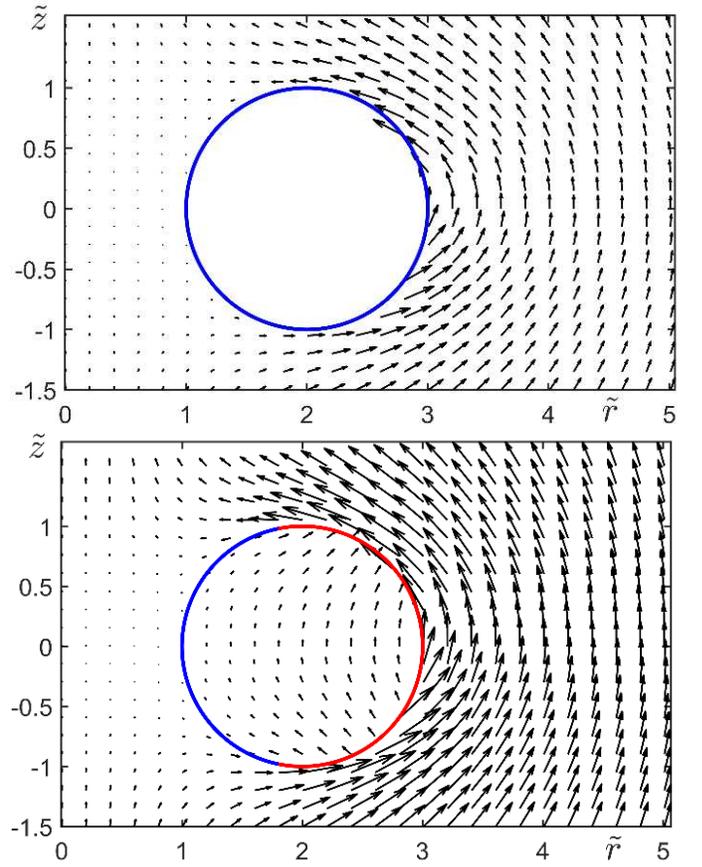

**Fig. 5.** Magnetic field near a toroidal shell. $\tilde{H}_x^e = 0.3$ (top) and 0.9 (bottom). Red line indicates the $|\tilde{J}| > 1$ zone.

TABLE II
Convergence of solutions for a spherical shell: Maximal absolute errors and computation times.

| $N$ | $\Delta \tilde{J}$ | $\Delta \tilde{E}$ | CPU time, sec. |
|---|---|---|---|
| 25 | 1.4e-2 | 1.9e-1 | 0.01 |
| 50 | 4.5e-3 | 4.6e-2 | 0.01 |
| 100 | 5.1e-4 | 7.9e-3 | 0.03 |
| 200 | 5.4e-5 | 3.4e-4 | 0.09 |
| 400 | 9.8e-7 | 1.6e-6 | 0.38 |
| 800 | --- | --- | 1.68 |

TABLE III
Convergence of solutions for a toroidal shell: Maximal absolute errors and computation times.

| $N$ | $\Delta \tilde{J}$ | $\Delta \tilde{E}$ | CPU time, sec. |
|---|---|---|---|
| 50 | 2.4e-2 | 2.8e-1 | 0.02 |
| 100 | 1.6e-2 | 1.5e-1 | 0.06 |
| 200 | 1.6e-3 | 2.6e-2 | 0.18 |
| 400 | 2.5e-4 | 4.7e-3 | 0.38 |
| 800 | 7.0e-6 | 6.3e-5 | 3.26 |
| 1600 | --- | --- | 22.1 |

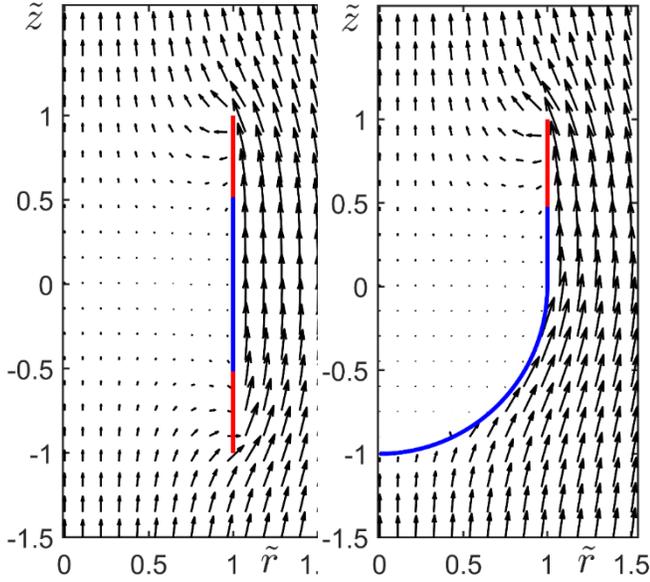

**Fig. 6.** The open and semi-closed cylindrical shells in the external field $\tilde{H}_z^e = 0.7$; the overcritical sheet current density zones are shown in red.

TABLE IV
| \multicolumn{4}{c}{Convergence of solutions for the semi-closed cylindrical shell.} |
| N | $\Delta \tilde{J}$ | $\Delta \tilde{E}$ | CPU time, sec. |
| --- | --- | --- | --- |
| 50 | 3.8e-4 | 3.4e-3 | 0.08 |
| 100 | 3.6e-5 | 3.0e-4 | 0.15 |
| 200 | 7.7e-6 | 7.4e-5 | 0.65 |
| 400 | 1.9e-6 | 1.8e-5 | 2.6 |
| 800 | 3.7e-7 | 3.6e-6 | 17 |
| 1600 | --- | --- | 93 |

## V. Conclusion

The proposed spectral method applies to axisymmetric problems involving shells of various shapes and accurately determines all relevant variables, including the electric field. We provided several examples, such as modeling magnetic shielding by a superconducting sphere. Numerical simulations showed very fast convergence and efficiency of this method. While the method is limited to axisymmetric problems, its high accuracy makes it useful in broader contexts: the solutions obtained can serve as benchmarks for other numerical methods when analytical solutions are unavailable, as in non-flat superconducting film problems.

## Appendix: Kernel Singularity Analysis

The singularity of $\mathcal{A} = \mathcal{K}(s,s')\chi(s')$ is due to $K(m) \sim -\ln(1-m)/2$ as $m \to 1$. Clearly, $m \leq 1$ and tends to 1 as $s' \to s$. We use the values $\mathcal{A}(s_i, s')$ of this kernel for $-1 \leq s' \leq 1$. Our set of interpolation knots, $\{s_i\}$, does not contain $\pm 1$, so $R(s_i) > 0$. Denoting $\delta = s - s'$ we obtain

$$1 - m = \frac{(dR(s)/ds)^2 + (dZ(s)/ds)^2}{4R(s)^2}\delta^2 + O(\delta^3) = \frac{\chi(s)^2}{4R(s)^2}\delta^2 + O(\delta^3),$$

$$(1-m/2)K(m) \sim \frac{1}{2}\left[-\frac{1}{2}\ln\frac{\delta^2}{R(s)^2}\right] \sim -\frac{\ln|\delta|}{2}.$$

$$\frac{1}{\sqrt{m}}\sqrt{\frac{R(s)}{R(s')}} = \frac{\sqrt{(R(s')+R(s))^2 + (Z(s)-Z(s'))^2}}{2R(s')} = 1 + O(\delta),$$

Hence, $-(\chi(s)/2)\ln|s-s'|$ can be taken as the singular part of $\mathcal{A}$. The regular part is, correspondingly,

$$\mathcal{U} = \mathcal{K}(s,s')\chi(s') + (\chi(s)/2)\ln|s-s'|.$$

This formula cannot be applied if $s = s'$. Using

$$(1-m/2)K(m) - E(m) = \frac{1}{2}\ln\left(\frac{4}{\sqrt{1-m}}\right) - 1 + O(1-m) =$$

$$\frac{1}{2}\ln\left(\frac{8R(s)}{|\delta|\chi(s)}\right) - 1 + O(\delta^2) =$$

$$\frac{1}{2}\ln\left(\frac{8R(s)}{\chi(s)}\right) - 1 - \frac{1}{2}\ln|\delta| + O(\delta^2)$$

we find $\mathcal{U}(s,s) = \chi(s)\left(\frac{1}{2}\ln\left[\frac{8R(s)}{\chi(s)}\right] - 1\right)$.